# Quantum path interferences in high-order harmonic generation from aligned diatomic molecules

S. Chatziathanasiou,[1,2] I. Liontos,[1] E. Skantzakis,[1] S. Kahaly,[3] M. Upadhyay Kahaly,[3] N. Tsatrafyllis,[1] O. Faucher,[4] B. Witzel,[5] N. Papadakis,[1] D. Charalambidis,[1,2,3] and P. Tzallas[1,3,*]

[1]*Foundation for Research and Technology–Hellas, Institute of Electronic Structure and Laser, P.O. Box 1527, GR-71110 Heraklion, Greece*
[2]*Department of Physics, University of Crete, P.O. Box 2208, GR-71003 Heraklion (Crete), Greece*
[3]*ELI–ALPS, ELI–Hu Kft., Dugonics tér 13, H-6720 Szeged, Hungary*
[4]*Laboratoire Interdisciplinaire CARNOT de Bourgogne, UMR 6303 CNRS-Université Bourgogne Franche-Comté, Boîte Postale 47870, 21078 Dijon, France*
[5]*Université Laval, Centre d'Optique, Photonique et Laser (COPL), Quebéc, Canada G1V 0A6*

Electron quantum path interferences in strongly laser–driven aligned molecules and their dependence on the molecular alignment is an essential open problem in strong–field molecular physics. Here, we demonstrate an approach which provides a direct access to the observation of these interference processes. The approach is based on the combination of the time–gated–ion–microscopy technique with a pump–probe arrangement used to align the molecules and generate high–harmonics. By spatially resolving the interference pattern produced by the spatiotemporal overlap of the harmonics emitted by the short– and long–electron quantum paths, we have succeeded in measuring *in-situ* their phase difference and disclose their dependence on molecular alignment. The findings constitute a vital step towards understanding of strong–field molecular physics and the development of attosecond spectroscopy approaches without the use of auxiliary atomic references. The manuscript is published in PHYSICAL REVIEW A **100**, 061404(R) (2019).

*Corresponding author: ptzallas@iesl.forth.gr

The process of high-harmonic generation (HHG) [1,2] in atoms or molecules is the foundation for the development of attosecond science [3-6]. It is governed by the recombination of localized electron wave packets ejected into the continuum and driven back toward the core upon reversal of the linearly polarized driving field (*three-step* model) [7, 8]. According to this model, for a given driving laser intensity, two quantum interfering electron wave packets, namely the Short (*S*) and the Long (*L*), with different flight times $\tau_q^L$ and $\tau_q^S$ (with $\tau_q^L > \tau_q^S$) contribute to the emission of plateau harmonics $q$ (with $\hbar\omega_q < IP + 3.17U_p$, where $IP$ is the ionization potential of the atom or molecule and $U_p$ the electron ponderomotive energy) with phases $\phi_q^{L,S} \approx -U_p\tau_q^{L,S}$. Additionally, it is well known [9, 10] (and references there in), that the harmonic beam generated by *L*-electron paths has larger divergence compared to the beam generated by the *S*-electron paths, while the relative contribution of the *S*– and *L*–electron paths in the outgoing from the medium harmonic beam can be controlled by the relative position of the focus of the driving laser field with respect to the harmonic generation medium.

Although the majority of the studies in attosecond science have been performed using atoms as HHG medium, over the last decade, HHG from molecules has attracted a keen interest from the scientific community as an important tool offering access to attosecond molecular spectroscopy and the ability of controlling the properties of the harmonics. This is because the properties of the emitted harmonics are strongly connected, via the electron recollision process, with the molecular structure. The majority of the studies in this research direction have been performed using aligned linear molecules in proof of principle experiments [11-20]. Such investigations are typically performed using field–free molecular alignment approaches based on pump–probe–type configurations (see for example ref. [21]) where the pump pulse is used to align the molecules and the probe to generate the high–harmonics. Towards attosecond molecular spectroscopy, semi–classical theories and advanced mathematical algorithms, have been combined with sophisticated experimental methods and calibration approaches [14] aiming to obtain the dependence of the structure of the molecular orbital on the amplitude and phase of the emitted harmonics considering that the harmonics emitted only from the *S*-electron quantum paths. The latter relies on the implementation of approaches like two–color (infrared (IR) and XUV) cross correlation [22, 23], transient molecular grating [24, 26], schemes based on two spatially separated XUV–sources [27–29], and gas mixtures [30–35]. However, a weakness in the majority of the demonstrated approaches is the use of auxiliary atomic references relying on the assumption that they have similar response to strong field as the molecule under investigation [14]. This restricts their applicability in revealing the dynamical information in a more versatile way and limits the accuracy of the measurements.

Recent theoretical calculations have shown that a way to overcome this obstacle and purify the results is to develop a direct measurement approach (a method which does not need the use of auxiliary atomic references) based on the use of S– and L–quantum path interferences in a molecular system [36]. However, such interferences were never observed experimentally in a molecular system and hence their applicability on attosecond spectroscopy remained an open question.

Here, by employing a direct measurement approach, we demonstrate the existence of the $S$– and $L$–electron quantum path interference effects in strongly laser driven molecules and we measure their dependence on the molecular alignment. Specifically, we show that the presence of S– and L–quantum paths in the harmonic generation is of significant importance for accessing intricacies associated with the dependence of the dynamics of the recollision process on the molecular structure. By exploiting the beneficial experimental simplification of using only a single XUV beam produced by the S– and L–electron quantum paths of aligned $N_2$ molecules, we show that the spatial intensity distribution, recorded in the region where the generated XUV beam is subsequently focused, carries information about the harmonic generation process during the alignment. This was achieved by employing a pump–probe arrangement (used to trigger the alignment process and generate the high-harmonics), in combination with a time-gated ion microscope [37-40]. The latter maps the spatial XUV intensity distribution onto a spatial ion distribution (produced in the XUV focal area through a single–XUV–photon ionization process of Argon (Ar) atoms). In this way, the dynamics of the alignment was obtained by measuring, the spatially integrated $Ar^+$ yield (produced along the propagation axis around the XUV focus and which is proportional to the XUV intensity) as a function of the pump–probe delay, while the dependence of the electron quantum path interference on the molecular orientation was revealed by spatially resolving the ion distribution along the propagation axis around the focus.

The experiment is performed at the Foundation for Research and Technology-Hellas (FORTH) utilizing a 20 TW 10 Hz Ti:Sapphire laser system which delivers ≈ 25 fs linearly polarized laser pulses with central wavelength at ≈ 800 nm. The experimental setup is shown in Fig. 1(a). The energy of the laser beam entering the setup was up to ≈ 40 mJ/pulse. The laser beam was divided in two, providing a sequence of pulses with adjustable delay. To improve the degree of alignment, the pump pulse was temporally stretched to ≈ 100 fs by inserting two BK7 plates of 2 cm total thickness. The probe pulse was fully compressed and used to generate the high–harmonics. Both pulses were focused by a 3 m focal length lens and crossed at a small angle (≈ 2 deg) in a pulsed jet filled with $N_2$ gas. The two foci were placed on the $N_2$ gas jet favoring the emission of S– and L–quantum path harmonics [41-43]. The intensity of the pump pulse was adjusted to have the optimum molecular alignment conditions ($I_\ell^{(pump)} \approx 5 \times 10^{13}$ W/cm$^2$), while for the probe was set just below the saturation intensity of the harmonic generation process ($I_\ell^{(probe)} \approx 10^{14}$ W/cm$^2$). After the $N_2$ jet, a Si plate was placed at Brewster's angle (for the fundamental wavelength) to separate the harmonics from the fundamental, which is transmitted while the harmonics are reflected towards the detection area. The XUV radiation, after reflection from the Si plate, passes through a 5 mm–diameter aperture (A) and a 150 nm–thick Sn filter (F) which transmits the harmonics from 11th to 15th (with relative amplitudes 0.6 (11th):1(13th):0.8(15th)) and blocks any residual part of the IR probe beam. Subsequently, the XUV beam was focused into the Ar gas jet by a spherical gold mirror (SM) of 5 cm focal length. Care was taken to fix the angle of incidence on the gold mirror at $\approx 0$ deg. The images were acquired by the transversely placed ion–imaging–detector (IM) (magnification of $\approx 100$) which records the spatial distribution of Ar$^+$ produced around the XUV focus (left panel of Fig. 1(b)). The contribution of the S– and L–quantum path harmonic emission is shown in the right panel of Fig. 1(b) which depicts the Ar$^+$ signal along the $y$–axis measured at a distance $z \approx 140$ $\mu$m ($z$ is the propagation axis of the XUV beam) before the focus. Because the harmonics used in the experiment are lying in the low photon energy region of the plateau harmonic spectrum it is considered as mainly generated by the HOMO orbital [44]. The contribution of the two–center interference effect [12, 13, 15] was neglected as the electron wavelength ($\lambda_e \approx 2.7$ Å) corresponding to the emission of the 11th–15th harmonics is larger than the $\approx 1.098$ Å internuclear distance of the N$_2$ molecule.

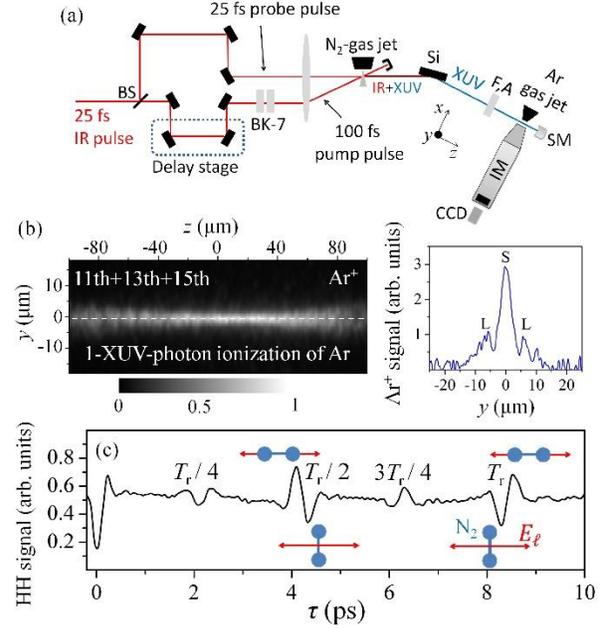

FIG. 1. (a) Experimental setup. (b) Left panel: $Ar^+$ distribution produced by single-photon ionization around the XUV focus. Eighty shots were accumulated for each image. The images are the projections of the three-dimensional ion distributions with cylindrical symmetry on the detection plane. $z = 0$ is the XUV focus position. Right panel: Far-field XUV beam profile showing the contribution of the $S$ and $L$ quantum path harmonics. (c) Harmonic signal as a function of the pump-probe delay $\tau$. $T_r$ is the rotational period of $N_2$. The alignment of the $N_2$ molecule with respect to the polarization of the probe field is depicted around $T_r$ and $T_r/2$.

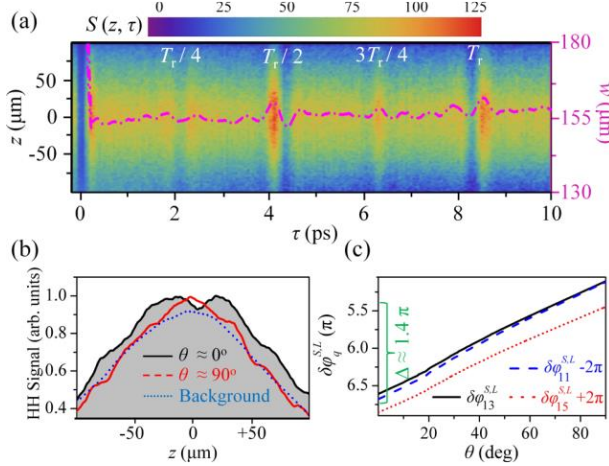

FIG. 2(a) Lineouts of the Ar$^+$ distribution at $x = 0$, $y = 0$, $z$ as a function of $\tau$. The pink dashed line shows the width $w$ of the lineouts as a function of $\tau$. (b) Normalized to unity, lineouts of the Ar $^+$ distribution at $x = 0$, $y = 0$, $z$ showing how the modulation of the width and the double– and single–peak structure resulted by the interference of $S$– and $L$–electron quantum paths around $T_r/2$, is built on the background signal. The gray–filled area and the red line correspond to the lineouts at $\tau \approx 4.1$ ps ($\theta \approx 0^o$) and $\tau \approx 4.3$ ps ($\theta \approx 90^o$), respectively. The blue–dotted line shows the background signal. (c) The calculated $\delta\phi_q^{S,L}(\theta)$.

At a first step, the evolution of the molecular alignment was traced by integrating the lineouts of the Ar$^+$ distribution at $x = 0$, $y = 0$, $z$ (white dashed line in the left panel of Fig. 1(b) for each pump–probe delay ($\tau$) Fig. (1c). In particular, the recorded trace shows the dependence of the XUV yield on the orientation of the molecular axis during the evolution of the field–free alignment process. In agreement with the findings of Refs. [11, 15, 45, 46] the harmonic yield depicts a modulation every alignment revival at $\tau \approx T_r/4$ (where $T_r \approx 8.4$ ps is the rotational period of N$^2$) i.e. at $\tau \approx 2.1$ ps, $\approx 4.2$ ps, $\approx 6.3$ ps, and $\tau \approx 8.4$ ps. The maximum (see e.g. at $\tau \approx 4.1$ ps and $\approx 8.5$ ps) and minimum (see e. g. at $\tau \approx 4.3$ ps and $\approx 8.3$ ps) values correspond to the times that the molecular axis is parallel and perpendicular to the polarization of the probe beam, respectively.

The dependence of the dynamics of the S– and L–quantum paths on the alignment of the molecular axis is expected to be imprinted in the spatial XUV intensity distribution along $z$–axis around the XUV focus. Figure 2(a) shows the lineouts of the Ar$^+$ distribution as a function of $\tau$. Besides the periodic modulation of the amplitude (shown in Fig. 1(c)), the width w of the lineouts (measured at the full–width of half maximum of the recorded signal) also depicts a clear modulation every $\tau \approx T_r/4$ (pink–dashed–dot line in Fig. 2(a). The latter is attributed to interference effects exhibited in the XUV intensity distribution pattern due to the spatiotemporal overlapping of the harmonics emitted by S– and L– interfering quantum paths as it was demonstrated in [37]. When the S– and L–quantum paths contribute to the harmonic emission, the width of the intensity distribution along the XUV focus depends on the phase difference $\delta\phi_q^{S,L} = \phi_q^S - \phi_q^L$, where $\phi_q^{S,L}$ is the phase of the $q^{th}$ harmonic emitted by the S– and L–quantum paths, respectively. As is shown in Ref. [37], for $\delta\phi_q^{S,L} = 2n\pi$ ($n = 0,1,2...$) the interference results in an intensity distribution having a single peak structure (narrow width) (red line in Fig. 2(b)) while in case of $\delta\phi_q^{S,L} = (2n + 1)\pi$ the distribution is broader having double peak structure (gray filled area in Fig. 2(b)).

It is instructive to note that in high–harmonic generation from atoms, $\delta\phi_q^{S,L} \approx -U_p\delta\tau_q^{S,L} = -(\delta\alpha_q^{S,L})I_\ell$. $U_p$, and $I_\ell$ are the ponderomotive potential and the intensity of the driving laser field, respectively. $\tau_q^{S,L}$ are the electron traveling times associated with the S– and L–quantum paths leading to the emission of the $q^{th}$ harmonic and for Ar atom $\delta\alpha_q^{S,L} \approx 20 \times 10^{-14}$ rad · cm$^2$/W [47] is a constant (which depends on the ionization potential (IP) of the atom) corresponding to the difference of the phase coefficient $\alpha_q^{S,L}$ of the $q^{th}$ harmonic. In this case $\delta\phi_q^{S,L}$ can be measured by varying $I_\ell$.

In an anisotropic system such as a linear molecule, $\delta\phi_q^{S,L}$ will also depend on the angle between the molecular axis and polarization of the driving field [48].

In the present work, because $I_\ell$ is constant, the change in w is attributed to the dependence of the $\alpha_q^{S,L}$ on the angle $\theta$ between the molecular axis and the polarization of the driving field i.e. $\delta\phi_q^{S,L}(\theta) \approx -U_p \delta\tau_q^{S,L}(\theta) = -(\delta\alpha_q^{S,L}(\theta))I_\ell$. Due to the dependence of the tunneling rate on the coupling $\mathbf{d}\cdot\mathbf{E}_\ell$ (where $\mathbf{d}$ is the induced transition dipole moment and $\mathbf{E}_\ell$ is the laser electric field) it is shown that the ionization rate (W) [44, 46, 49, 50] and the efficiency of the HH generation process [14, 45, 51] for $N_2$ in HOMO orbital drops by a factor of $\sim 6$ when the orientation of the $N_2$ axis changes from parallel to perpendicular with respect to the polarization of the driving field.

In the light of the above-mentioned arguments, we attempt to calculate $\delta\phi_q^{S,L}(\theta)$. Firstly we have used the dependence of W on $\theta$ calculated by the molecular tunneling ionization (MO-ADK) theory [52] in Refs. [44, 49] for $N_2$ in HOMO orbital. Then, we have solved the semi–classical three–step model [8] for Ar atom in an intensity range (from $I_\ell^{(max)} \approx 10^{14}$ W/cm$^2$ to $I_\ell^{(min)} \approx 8\times 10^{13}$ W/cm$^2$) which provides ionization rates that are matching those given by the MO–ADK theory i.e. $W(I_\ell^{(max)})/W(I_\ell^{(min)}) \approx 6$. Argon was selected as it is very similar to $N_2$ in its response to the strong laser field having almost the same IP ($IP_{Ar} = 15.76$ eV and $IP_{N_2} = 15.58$ eV) and intensity dependent ionization probability [53, 54]. We note that Ar is used only to facilitate the theoretical calculations in order to show that the effect we measure is compatible with established theory, and does not play the role of an auxiliary experimental reference. Fig. 2(c) shows the calculated $\delta\phi_q^{S,L}(\theta)$ for $q = 11,13,15$. It is evident that when the molecular axis changes from $\theta = 0$ deg to 90 deg with respect to the laser polarization, the value of $\delta\phi_q^{S,L}$ depicts a shift of $\Delta \approx 1.4\pi$. $\Delta = \delta\phi_\parallel^{S_\parallel,L_\parallel} - \delta\phi_\perp^{S_\perp,L_\perp}$, $\delta\phi_\parallel^{S_\parallel,L_\parallel} = \langle\delta\phi_\parallel^{S_\parallel,L_\parallel}(\theta=0)\rangle$, $\delta\phi_\perp^{S_\perp,L_\perp} = \langle\delta\phi_\perp^{S_\perp,L_\perp}(\theta=90)\rangle$, $\langle\delta\phi^{S,L}(\theta)\rangle$ is the phase difference corresponding to the carrier frequency of the XUV radiation, and $S_\parallel$, $L_\parallel$, $S_\perp$, $L_\perp$ are the electron quantum paths for $\theta = 0$ deg ($\parallel$) and 90 deg ($\perp$), respectively. Since the detected signal results from a harmonic superposition, hereafter, the term $\delta\phi_q^{S,L}$ is replaced by $\langle\delta\phi^{S,L}\rangle$. The fact that the calculated value of $\Delta$ exceeds the value of $\pi$ supports the presence of quantum path interference effects which are expected to be more pronounced around the revival times.

The existence of $S$– and $L$–electron quantum path interference effects is undoubtedly exhibited by the modulation width and the structure of the recorded pattern shown in Figs. 2(a, b), respectively. However, the interference fringe pattern cannot be clearly resolved in the recorded raw data of Fig. 2(a) as it is built on a large background as is shown in Fig. 2(b). The background signal [blue dashed line in Fig. 2(b)] is attributed to the influence of several factors such as, the unbalanced contribution between the $S$– and $L$–signal amplitudes, a difference in the focal position between $S$– and $L$–quantum paths [55] as well as in the inherently limited degree of molecular alignment. In order to better visualize the interference pattern in the measured data of Fig. 2(a), we have subtracted from each lineout of the raw data a normalized to the harmonic yield ($Y(\tau)$) background signal ($S_{BKG}(z,\tau_{random})$) obtained by averaging the lineouts in the region of 0.5 ps $< \tau_{random} <$ 1 ps where the molecules are nearly randomly oriented [blue–dotted line in Fig. 2(b)]. For the subtraction we have used the equation $S'(z,\tau) = S(z,\tau) - S_{BKG}(z,\tau_{random})\cdot Y(\tau)/\langle Y(\tau_{random})\rangle$, where $\langle Y(\tau_{random})\rangle$ is the average harmonic yield in the region of $\tau_{random}$. For simplicity, the coordinates $x = 0$ and $y = 0$ are omitted from the equation. The resulted signal $S'(z,\tau)$ is shown in Fig. 3(a).

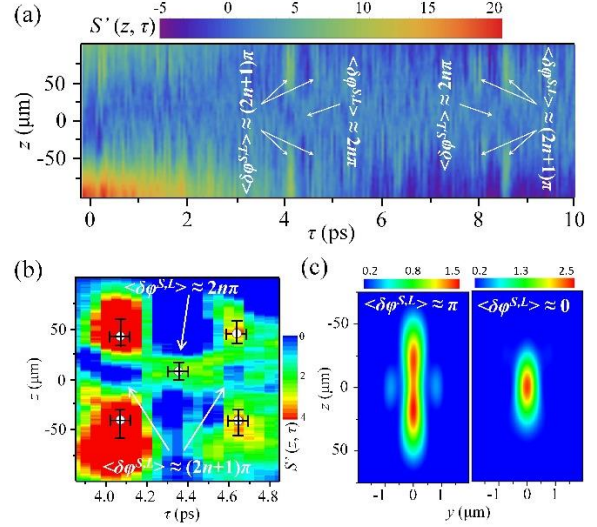

FIG. 3. (a) Lineouts of the Ar$^+$ distribution as a function of $\tau$ after the background subtraction. The white arrows depict the distinct areas around $T_r$ and $T_r/2$ where the double- and single peak structure along the $z$ axis is evident. (b) Lineouts of the Ar$^+$ distribution, around $T_r/2$. The filled white circles depict the position of the double- and single-peak structure that can be safely deduced from the experiment. The error bars represent one standard deviation of the mean. (c) Calculated XUV interference patterns around the focus for $\langle\delta\phi^{S,L}\rangle \approx \pi$ (left panel) and $\langle\delta\phi^{S,L}\rangle \approx 0$ (right panel). The theoretical calculations have been performed by considering that all the molecules are perfectly aligned along a direction defined by $\vartheta$ [shown in Fig. 2(c)].

A clear single and double peak structure with respect to $z \approx 0$, corresponding to a phase difference of $\langle\delta\phi^{S,L}\rangle \approx 2n\pi$ and $\langle\delta\phi^{S,L}\rangle \approx (2n+1)\pi$, respectively, appeared in the regions of $\tau \approx T_r/2$ and $\tau \approx T_r$. In the regions where the molecules are randomly oriented no pronounced structure was observed. We note, that although the followed background subtraction process fails to entirely subtract the background signal in the time delay region of <2 ps (which is out of interest in the present work) if there was any interference structure with strength similar to those appearing at $\tau \approx T_r/2$ and $\tau \approx T_r$, it would be possible to be resolved. For the $I_\ell^{(\text{probe})}$ used in the present experiment, the double and single peak structure appeared at $\tau_{\text{exp}}^{(\text{double})} \approx 4.06 \pm 0.05$ ps and $\tau_{\text{exp}}^{(\text{single})} \approx 4.35 \pm 0.05$ ps, respectively [Fig. 3(b)]. These values are very close to $\tau = 4.1$ ps and 4.3 ps for which the squeezing of the angular distribution of the molecules around $\theta = 0$ deg and 90 deg, respectively, is maximum, as shown by comparing Figs. 1(c), 2(a) and 3(a). These have been also confirmed by the analysis of the intensity distribution interference pattern calculated in the region around $\tau \approx T_r$. Figure 3(c) shows the calculated intensity distribution at the focus for $\langle \delta\phi^{S,L} \rangle \approx \pi$ (left panel) and $\langle \delta\phi^{S,L} \rangle \approx 0$ (right panel). The distribution was obtained by calculating the images of the focused harmonic beam with the phases $\phi_q^{S,L}$ used to obtain Fig. 2(c), the measured relative harmonic amplitudes, and the measured $S$– and $L$–quantum path harmonic beam profile shown in the right–panel of Fig. 1(b). We note that [besides a marginal constant Gouy phase shift which is included in the calculations of Fig. 3(c)] the spatiotemporal coupling effects in the harmonic generation area [10, 55] have negligible dependence on $\theta$ and thus have been neglected from the analysis. The images were calculated by the Debye integral [37, 56] after applying the Huygens–Fresnel principle on a spherical mirror with a 10 cm radius of curvature. In these calculations the appearance of the double and single peak structure is found to be at $\tau_{\text{cal}}^{(\text{double})} \approx 4.13$ ps and $\tau_{\text{cal}}^{(\text{single})} \approx 4.3$ ps, respectively, which is in fair agreement with the experimental values. The deviation observed between $\tau_{\text{exp}}$ and $\tau_{\text{cal}}$, might originate from the contribution of high–harmonics generated by other molecular orbitals (like HOMO–1, HOMO–2) [36]. However, this matter requires more elaborative considerations of multi-electronic molecular structure which is out of the scope of the present work.

In conclusion, by utilizing a direct measurement approach we have measured electron quantum path interference effects in strongly laser driven aligned $N_2$ molecules and quantified their dependence on the orientation of the molecule. It is found that the phase difference between the $S$– and $L$–quantum paths is shifted by a value larger than $\pi$ when the molecular axis is changing from parallel to perpendicular with respect to the polarization of the laser field. These findings, besides the fundamental interest on carrying out investigations in aligned diatomic molecules, in conjunction with longer wavelength–driven molecular–HHG schemes and coincidence detection arrangements, can be used for the development of self–referenced attosecond spectroscopy approaches [36, 57] and the recently demonstrated laser driven electron diffraction schemes [58, 59].

This work was supported by the: LASERLAB-EUROPE (Nr.: GA654168), NFFA–Europe (Nr.: 654360), "HELLAS-CH" (MIS 5002735) funded by the Operational Program (NSRF $2014 - 2020$) and co-financed by Greece and the European Union, the "GAICPEU" project funded by the Hellenic Foundation for Research and Innovation (HFRI) and the General Secretariat for Research and Technology (GSRT) (Nr.: 645), the IKY doctorate scholarship that is co-financed by Greece and the European Union (European Social Fund- ESF) through the Operational Programme «Human Resources Development, Education and Lifelong Learning 2014–2020, the CNRS and the EIPHI Graduate School (contract "ANR-17-EURE-0002"). ELI–ALPS is supported by the European Union and co–financed by the European Regional Development Fund (GINOP–2.3.6–15–2015–00001).

S. C., I. L., E. S. equally contributed to the experiment and data analysis; S. K. contributed in the experiment and the theoretical calculations; M. U. K., N. T performed the theoretical calculations; O. F. developed the pump-probe arrangement, contributed to data analysis and manuscript preparation; N. P. developed the data acquisition system; B. W., D. C. contributed to data analysis and manuscript preparation; P. T. conceived the idea, supervised the project, wrote the manuscript and contributed to all aspects of the work.